\documentstyle[12pt,epsfig]{article}
\begin{document}
\centerline{\Large\bf Reply to ``Comment on Fine Structure Constant}
\centerline{\Large\bf in the Spacetime of a Cosmic String''}
\vspace*{0.050truein}
\centerline{Forough Nasseri\footnote{Email: nasseri@fastmail.fm}}
\centerline{\it Physics Department, Sabzevar University of Tarbiat
Moallem, P.O.Box 397, Sabzevar, Iran}
\begin{center}
(\today)
\end{center}
\begin{abstract}
In this Reply, using E.R. Bezerra de Mello's comment, I correct
calculations and results presented in Phys. Lett. B 614 (2005) 140-142
about fine structure constant in the spacetime of a cosmic string.
\end{abstract}
In a recent letter \cite{1}, I analysed the Bohr's atom in the spacetime
of a cosmic string. I concluded that in the presence of a
cosmic string the fine structure constant reduces by a factor
$\pi^2 \epsilon_0 G \mu$, or as numerical results by a
factor $8.736 \times 10^{-17}$, see Eqs.(18) and (19) of \cite{1}.

Based on E.R. Bezerra de Mello's comment \cite{2}, the conclusions
presented in \cite{1} about the fine structure constant in the spacetime
of a cosmic string are not completely correct in the sence that the
results presented in \cite{1} are valid only in a special situation.
As explained in \cite{2}, the conclusions presented in \cite{1} are
correct if we assume as ideal not only the proton is placed on the cosmic
string but also the electron and the proton are in the plane orthogonal to
the cosmic string lying along the $z$-axis.

The purpose of this Reply is to correct calcultions presented in
\cite{1} in accordance with the observation made in \cite{2} and also
to present the correct value of the fine structure constant in the
spacetime of a cosmic string. As pointed out in \cite{2}, our approach
is correct only if we consider the proton and the electron placed
in a plane perpendicular to the cosmic string which lies along the
$z$-axis, with the proton on the cosmic string.

It should be emphasized that the mistake concerning the value of the
fine structure constant is due to the fact that we have used two
different unit systems which we corrected in this Reply.

Linet in \cite{4} has shown that the electrostatic
field of a charged particle is distorted by the cosmic string.
For a test charged particle in the presence of a cosmic string
the electrostatic self-force is repulsive and is perpendicular to the
cosmic string lying along the $z$-axis\footnote{Linet
in \cite{4} has used the mks units and in Eqs.(15) and (16) of \cite{4}
has obtained $f^z=f^{\phi}=0$ and
$$
f^{\rho} \sim \left( \frac{2.5}{\pi} \right)
\left( \frac{G \mu}{c^2} \right)
\left( \frac{q^2}{4 \pi \epsilon_0 \rho_0^2} \right)
$$
when $\mu \to 0$. Indeed we can put the fraction $\frac{2.5}{\pi}$
to be approximately equal to $\frac{\pi}{4}$. With this substitution
we obtain (\ref{11}) of this Reply.}
\begin{equation}
\label{11}
f^{\rho}\simeq\frac{\pi}{4} \frac{G \mu}{c^2} \frac{e^2}{4 \pi \epsilon_0 \rho_0^2},
\end{equation}
where $f^{\rho}$ is the component of the electrostatic self-forcs
along the $\rho$-axis in cylindrical coordinates and $\rho_0$ is the
distance between the electron and the cosmic string.

For the Bohr's atom in the absence of a cosmic string, the electrostatic
force between an electron and a proton is given by Coulomb
law\footnote{In fact in \cite{1} I have used the same letter $r$ to
define the distance from the electron to the cosmic string, and from
the electron to the proton. The force in Eq.(11) of \cite{1}
is perpendicular to the string, so we may use $\rho=\sqrt{x^2+y^2}$.
As to the force in Eq.(12) of \cite{1}, it is radial:
$r=\sqrt{x^2+y^2+x^2}$. I thank E.R. Bezzera de Mello for pointing
this out to me.}
\begin{equation}
\label{12}
\vec F=\frac{-e^2}{4 \pi \epsilon_0 r^2} \hat{r}.
\end{equation}

As discussed in \cite{2},
to obtain the fine structure constant in the spacetime of a cosmic
string we assume that the proton located on the cosmic string
lying along the $z$-axis. We also assume that the proton
located in the origin of the cylindrical coordinates and
the electron located at $\rho=\rho_0$, $z=0$ and $\phi=0$.
This means that the electron and the proton are in the plane
orthogonal to the cosmic string.

The orbital speed of the electron in the first Bohr orbit is 
\begin{equation}
\label{13}
v_1=\frac{e^2}{4 \pi \epsilon_0 \hbar}.
\end{equation}
The ratio of this speed to the speed of light, $v_1/c$, is
known as the fine structute constant
\begin{equation}
\label{14}
\alpha=\frac{e^2}{4 \pi \epsilon_0 \hbar c}.
\end{equation}

To calculate the fine structure constant in the spacetime of a cosmic
string we consider a Bohr's atom in the presence of a cosmic string.
For a Bohr's atom in the spacetime of a cosmic string, we
take into account in Eq.(\ref{11}) the sum of two forces, i.e.
the electrostatice force for Bohr's atom in the absence of a cosmic
string, given by Eq.(\ref{12}), plus the electrostatic self-force of
the electron in the presence of a cosmic string.
Because we assume that the proton located at the origin of the
cylindrical coordinates and on the cosmic string and also
the plane of electron and proton is perpendicular to the cosmic string
lying along the $z$-axis, the induced electrostatic self-force and the
Coulomb force are at the same direction, i.e. the direction of the
$\rho$-axis in cylindrical coordinates. Therefore, we can sum
these two forces
\begin{equation}
\label{15}
\vec F_{\rm {tot}}= \left( -\frac{e^2}{4 \pi \epsilon_0 \rho_0^2}+\frac{\pi}{4}
\frac{G \mu}{c^2} \frac{e^2}{4 \pi \epsilon_0 \rho_0^2} \right) \hat \rho.
\end{equation}
It can be easily shown that this force has negative value and
is an attractive force ($\frac{\pi G \mu}{4 c^2} < 1$).

The numerical value of the fine structure constant in the spacetime of
a cosmic string can be computed by Eq.(\ref{15}). The orbital
speed of the electron in the first Bohr orbit in the spacetime of a
cosmic string has positive value and is given by
\begin{equation}
\label{16}
{\hat v}_1=\frac{e^2}{4 \pi \epsilon_0 \hbar}
-\frac{\pi}{4} \frac{G \mu}{c^2} \frac{e^2}{4 \pi \epsilon_0 \hbar}.
\end{equation}
The ratio of this speed to the speed of light, ${\hat v}_1/c$,
is presented by the symbol $\hat \alpha$ which is the fine structure
constant in the spacetime of a cosmic string
\begin{equation}
\label{17}
\hat \alpha=\frac{e^2}{4 \pi \epsilon_0 \hbar c}
-\frac{\pi}{4} \frac{G \mu}{c^2} \frac{e^2}{4 \pi \epsilon_0 \hbar c}.
\end{equation}
From (\ref{14}) and (\ref{17}) we obtain
\begin{equation}
\label{18}
\frac{\alpha}{\hat \alpha}=\frac{1}{1-\frac{\pi G \mu}{4 c^2}}.
\end{equation}
This means that the presence
of a cosmic string causes the value of the fine structure constant
reduces by a factor $\frac{\pi G \mu}{4 c^2}$.
In the limit $\frac{G \mu}{c^2} \to 0$, i.e. in the absence of a cosmic
string, ${\alpha}/{\hat \alpha} \to 1$.
From Eq.(\ref{18}) we obtain
$(\alpha -\hat \alpha)/\alpha= \frac{\pi G \mu}{4 c^2}$.

The dimensionless parameter $\frac{G \mu}{c^2}$ plays an important role in the
physics of cosmic strings. In the weak-field approximation
$\frac{G\mu}{c^2} \ll 1$.
The string scenario for galaxy formation requires $\frac{G \mu}{c^2} \sim 10^{-6}$
while observations constrain $\frac{G \mu}{c^2}$ to be less
than $10^{-5}$. For more details about the range of value of
$\frac{G \mu}{c^2}$ see \cite{5}.
Inserting $\frac{G\mu}{c^2} \sim 10^{-6}$ in the right-hand side of
Eq.(\ref{18}) yields
\begin{equation}
\label{19}
\hat \alpha \sim
\left( 1-\frac{\pi}{4} \times 10^{-6} \right) \alpha.
\end{equation}
{\bf Acknowledgments:}\\
It is a pleasure to thank E.R. Bezerra de Mello,
N. Pileroudi and S.A. Alavi for
helpful comments. I thank Amir and Shahrokh for useful helps.

\end{document}